\documentclass[letterpaper]{article} 
\usepackage[]{aaai25}  
\usepackage{times}  
\usepackage{helvet}  
\usepackage{courier}  
\usepackage[hyphens]{url}  
\usepackage{graphicx} 
\usepackage{natbib}  
\usepackage{caption} 
\usepackage{algorithm}
\usepackage{algorithmic}

\newcommand{\answerYes}[1]{\textcolor{blue}{#1}} 
\newcommand{\answerNo}[1]{\textcolor{teal}{#1}} 
\newcommand{\answerNA}[1]{\textcolor{gray}{#1}} 
 
\usepackage{newfloat}
\usepackage{listings}
\floatstyle{ruled}
\newfloat{listing}{tb}{lst}{}
\floatname{listing}{Listing}
\usepackage{algorithm}
\usepackage{algorithmic}
\usepackage{float}

\usepackage{multirow}
\usepackage{enumitem}
\usepackage{comment}
\usepackage{mdframed}
\usepackage{amsmath}
\usepackage{breakurl}
\newlist{questions}{enumerate}{2}
\setlist[questions,1]{label=RQ\arabic*.,ref=RQ\arabic*}
\setlist[questions,2]{label=(\alph*),ref=\thequestionsi(\alph*)}

\newcommand{\changed}[1]{\textcolor{black}{#1}}

\newcommand{\g}{\cellcolor{lightgray!50!white!50}}

\usepackage{newfloat}
\usepackage{listings}
\DeclareCaptionStyle{ruled}{labelfont=normalfont,labelsep=colon,strut=off} 
\floatstyle{ruled}
\newfloat{listing}{tb}{lst}{}
\floatname{listing}{Listing}
\title{Exploring YouTube's Political Communication Networks\\during the 2024 French Elections}
\author{
Caroline Violot, Vera Sosnovik, Mathias Humbert
}
\affiliations{University of Lausanne\\
\{caroline.violot, vera.sosnovik, mathias.humbert\}@unil.ch
}

\usepackage[table]{xcolor}
\begin{document}
\maketitle

\begin{abstract}
	In 2024, France was shaken by the far-right National Rally's victory in the European elections. In response to this unprecedented result, French President Emmanuel Macron dissolved the National Assembly, triggering legislative elections just two weeks later. A whirlwind campaign followed, partly on social media, as is now the norm, and concluded with the victory of a left-wing coalition. 
	This article examines the YouTube activity of two key actors during this period---news media and politicians---and the commenting behavior they generated. We built a dataset of 35 news media channels, 28 politicians \& parties channels, 43.5k videos posted from three months before the European elections to one week after the second round of the legislative elections, and 7.4M associated comments. We examined upload activity and engagement across political orientations and used network analysis methods to uncover the structure of their commenting communities. We also identified politicians’ appearances on news media channels and assessed their impact on commenting user bases. 
	Our findings show that, among politicians and parties channels, far-right and left-wing ones were significantly more active and received substantially higher engagement (views, likes, and comments) than other groups, with denser and more clustered commenting communities. About 7\% of commenters commented across political orientations and were much more active than in-group commenters. News media channels tended to favor politically aligned guests, while centrist politicians were overrepresented. Finally, politicians’ presence in the videos of a specific news media channel increased the share of commenters who were active on this channel and political channels, regardless of their orientation.
\end{abstract}

%

\section{Introduction}

Over the past two decades, social media have become a central component of political communication, gradually supplementing traditional means such as radio, television and press interviews, and campaign rallies~\cite{SeverinNielsen_2023}. Social media allow direct communication between politicians and the public, bypassing journalistic mediation~\cite{Chadwick_2017}, and enable two-way interaction as viewers can respond to and share politicians' content~\cite{Peeters_Opgenhaffen_Kreutz_VanAelst_2023}. \changed{Populist movements, including far-right ones, have especially flourished online, fostering alleged ties with “the people” and amplifying cultural threat narratives through anecdotal sharing~\cite{Engesser_Ernst_Esser_Büchel_2017}.}
News media also increasingly use social media to promote content and engage audiences~\cite{Rajapaksha_Farahbakhsh_Crespi_2024}, alongside their own communication channels, such as TV, radio, and print. \changed{Despite leveraging social media to circumvent mainstream news outlets, frequently condemned by populists politicians as part of a corrupt elite~\cite{Engesser_Ernst_Esser_Büchel_2017},} many politicians, including far-right ones, still re-purpose clips from news media appearances, content also shared by news media outlets on their own social channels. This shows a complex relationship between politicians, news media, and social media.

\changed{One aspect of this relationship unrolls} on YouTube where both politicians' videos and news media videos featuring politicians attract much interest and are highly commented by users, supporters or opponents~\cite{Wu_Resnick_2021}. YouTube comments have been part of the platform since its early days and a majority of users reported systematically reading the first two or three comments~\cite{Schultes_Dorner_Lehner_2013}. As a result, comments are a key feature of the platform, enabling significant influence with minimal effort. Active commenters represent only a small minority, and tend to hold more extreme views than the ``silent majority''~\cite{Bail_2022}. Moreover, it was repeatedly shown that comments influence readers in their perception of a video~\cite{Walther_DeAndrea_Kim_Anthony_2010,Hsueh_Yogeeswaran_Malinen_2015,Searles_Spencer_Duru_2020}. The use of social media by politicians and the engagement they generate have been previously studied in a variety of contexts and from different research perspectives, from computer science to political science~\cite{Rajapaksha_Farahbakhsh_Crespi_2024, Boulianne_Larsson_2023, Peng_2021, Moller_Kuhne_Baumgartner_Peter_2019, Engesser_Ernst_Esser_Büchel_2017}. Likewise, relationships between user bases of news media accounts of diverse political leanings were previously explored~\cite{Wu_Resnick_2021, Cage_Herve_Mazoyer_2020, Cointet_Cardon_Mogoutov_OogheTabanou_Plique_Morales_2021}. However, the vast majority of previous research focuses on the US context, which is bipartite, whereas the French political landscape is multiparty, with the last two presidential runner-ups not belonging to traditional left-wing or right-wing parties. 

In this study, we broaden the scope of analysis to include both politicians/political parties channels \changed{(later referred to as PP channels)}, and news media channels \changed{(later referred to as NM channels)}. We ground our work in a highly intense election period in France, during which two distinct elections were held within four weeks, comprising a total of three rounds of voting, one of which concluded with the victory of a far-right party.  \changed{We chose to focus on the YouTube platform as 61\% of French people use it weekly, with 56\% of them turning to it specifically to get informed~\cite{youtubestats}}. We also account for recent developments on the YouTube platform, particularly the introduction of short-form videos in 2021 under the term \textit{Shorts}~\cite{YouTubeShortsIntro}, and observe the diverse and unevenly successful uses of this new format. \changed{To better characterize this unprecedented situation, we further compare our results with previous elections data.}

This work is organized around two main axes of interest. First, we explore how NM and PP channels use the YouTube platform during the French European and legislative elections, as well as the reception from their audiences. To this end, we provide a novel characterization of channels within the French political-media ecosystem, encompassing both NM and PP YouTube channels. We analyze their uploading activity and the structure of their commenting user base, highlighting differences based on channel type and political orientation. The second axis of research focuses on the relationship between NM and PP channels, particularly the overlap of their respective commenting audiences, and the editorial choice made by NM channels regarding which political guests are shown on YouTube. We divide these two axes of investigation into the four research questions below.
\begin{enumerate}[leftmargin=1.5\parindent,align=left,labelwidth=\parindent]
	\item[\textbf{RQ1}] What characterizes NM and PP channels in terms of uploading behavior, and how can it be linked to the political orientation of the channels?\label{RQ:1}
	\item[\textbf{RQ2}] How do audience engagement patterns (views, likes, and commenters’ behavior) differ between NM and PP channels?\label{RQ:2}
	\item[\textbf{RQ3}] What are the relationships between the commenting user bases of PP channels, and how do these relate to those of NM channels?\label{RQ:3}
	\item[\textbf{RQ4}] Which politicians appear on NM channels, and how does this affect the commenting user base?\label{RQ:4}
\end{enumerate} 

To explore these questions, we analyze YouTube channels of the main French news outlets and the main French politicians and parties. We collected channels information, videos metadata and comments of 35 NM channels and 29 PP channels, over a four-month period that included the French European elections and the subsequent legislative elections, resulting in a dataset of 43.5k videos, and 7.4M comments. \changed{We also collected videos of the same channels posted during previous French legislative elections, in June 2017, obtaining 28.4k additional videos, to observe the evolution of channels' practices over the years.} We describe channels by analyzing their uploads and engagement metrics, and their commenting communities using three different graph representations: a bipartite videos-commenters graph, a commenters-projected graph, and a novel combination of the two. We also used pre-trained LLMs to investigate the presence of politicians in NM channel videos and how this affected the overall commenting network. 

Our results highlight a different usage of YouTube, and a different reception from its audience, according to political orientation and channel type (NM versus PP). We show a conditional effect of Shorts, benefiting only channels with an already strong user base, and that PP left-wing and far-right channels attract more engagement. Additionally, our network analysis shows that PP left-wing and far-right channels also foster tighter, more clustered communities of commenters. We also evaluate the proportion of commenters interacting with both NM and PP channels, and/or commenting on politically opposite channels, and find a core of very active, cross-commenting users. Finally, \changed{we reveal that far-right politicians were increasingly featured in center- and right-leaning NM channels and} that when politicians were discussed or featured in NM channel videos, the overlap between the NM channel's commenters and commenters from PP channels increased significantly, regardless of the orientation of the politicians. 

\section{Related Work}\label{sec:related}

\paragraph{Social network analysis approaches.}


At the early stages of YouTube,  \citet{Cheng_Dale_Liu_2008} provided a large-scale analysis of the platform, using data from over 3 million videos. 
They explored the YouTube items' statistical properties, such as videos' length, engagement metrics evolutions, and popularity distribution, and its network characteristics, showing the small-world characteristics~\cite{Travers_Milgram_1969} exhibited by the platform, extending results of \citet{Mislove_Marcon_Gummadi_Druschel_Bhattacharjee_2007} on online social networks to YouTube. 
\citet{Wattenhofer_Wattenhofer_Zhu_2012} had access to the complete YouTube subscription graph and comment graph. Using graph-based methods, they examined social interaction and popularity on the platform and compared them with traditional online social networks of the time. 
They revealed that popularity (number of subscribers) was better correlated with the popularity of one’s most successful content rather than with the summary measures of content popularity. 

Previous work on commenters' influence includes \citet{OCallaghan_Harrigan_Carthy_Cunningham_2012}, which constructed a co-commenter network based on comment similarity and detected spam bots using network analysis methods. 
They discovered that these bots' comments were frequently linked to coordinated campaigns targeting multiple videos and spanned over extended periods. The collective influence of groups of commenters was also analyzed by \citet{Alassad_Agarwal_Hussain_2019}, which proposed a new method to identify coordinated commenters' behavior and showed that these coordinating groups have significantly higher interacting/influencing power than other communities.
\citet{Shajari_Alassad_Agarwal_2024} categorized channels based on the commenting patterns they exhibited and revealed common patterns associated with suspicious behavior. 
\citet{Byun_Jang_Baek_2023} showed how video engagement was impacted by a video creator commenting their video or answering comments, using a hierarchical regression analysis. 

Cross-commenting and in-group commenting patterns were explored, especially on Twitter/X, until recently, where ~\citet{Cetinkaya_Ghafouri_SuarezTangil_Such_Elmas_2025} showed that cross-partisan interactions persist on Twitter and vary by topic, revealing where toxicity concentrates. Focusing on YouTube, \citet{Wu_Resnick_2021} analyzed comments posted under US news channels, evaluated the political partisanship of commenters, and showed that cross-commenters were not unusual but were more toxic than in-group commenters. Focusing on French media websites, \citet{Cointet_Cardon_Mogoutov_OogheTabanou_Plique_Morales_2021} identified structural features of their hyperlinks network and concluded that the French media ecosystem did not suffer from the same level of polarization as in the US.

\paragraph{Elections and social media commenters.}

Commenters' behavior in the context of elections has already motivated a large body of research. \citet{Nizzoli_Tardelli_Avvenuti_Cresci_Tesconi_2021} uncovered coordinated behavior on Twitter around the 2019 UK General Election 2021.
Covering the same election, and also on Twitter, \citet{Hristakieva_Cresci_DaSanMartino_Conti_Nakov_2022} proposed an alternative method to discover propaganda operation and coordinated behavior, and evaluated its impact and influence.
\citet{Peeters_Opgenhaffen_Kreutz_VanAelst_2023} studied the engagement Belgium politicians' posts received on Facebook around elections. 
\citet{Chen_Wang_2022} analyzed comments posted on YouTube under misinformation-based political advertisements around the 2020 US election. \citet{Flamino_Galeazzi_Feldman_Macy_Cross_Zhou_Serafino_Bovet_Makse_Szymanski_2023} compared the volume of biased political content between two US presidential elections (2016 and 2020) and observed an increase in online polarization. Finally, \citet{Sosnovik_Violot_Humbert_2025} also focused on the 2024 French elections period and conducted an in-depth analysis of the different topics discussed by the political and news media actors on YouTube during that period.

Contrary to previous work on YouTube comments around election times, we focus on network dynamics of commenters rather than on comment content, and we incorporate both NM and PP channels altogether, in the French context. 

\section{Data Collection and Processing}\label{sec:data_collection}

\paragraph{News media channels choice.}
We collected a list from the CSA\footnote{The Conseil Supérieur de l'Audiovisuel (CSA), now Arcom, is the French authority in charge of regulating TV and radio channels.} of  TV and radio channels that invite politicians and cover political news~\cite{CSA_TV_list}, to which we added the main French daily newspapers and weekly magazines that cover political and societal news from a list produced by the ACPM~\cite{ACPM_list}.\footnote{Alliance pour les Chiffres de la Presse et des Médias.}  Finally, we added two ``pure players'', Blast and Le Média, which have a strong presence online. Not all NM were present on YouTube, and we obtained a final list of 35 channels. 

\paragraph{Politicians and parties channels choice.} 

\changed{We identified 13 parties using the list of the 2024 European election results~\cite{electionsEuropeenes} and the list of legislative campaign candidates~\cite{list_legislative}. To this list, we added the heads of parties and the heads of European lists. We then searched for active YouTube channels of identified politicians and parties, and obtained 29 channels in total.}

\paragraph{Grouping the channels.}

In some parts of the analysis, for ease of comprehension and visualization, we regrouped channels based on their political orientation. For PP channels, we regrouped all channels belonging to the same party together, and further grouped parties by their political orientation---far-left, left, centrist, right, and far-right---as given by the \textit{Ministry of the Interior and Overseas Territories}.\footnote{www.legifrance.gouv.fr/circulaire/id/45472} 
For NM channels the regrouping was more challenging, considering the fact that political bias attribution is a complicated and subjective task. We started our classification by referencing Media Bias/Fact Check\footnote{mediabiasfactcheck.com/} and Eurotopics\footnote{eurotopics.net/}, using the simplified categories \textit{right}, \textit{center}, and \textit{left}. Some outlets were not mentioned in our sources, and some had a recent change in ownership or editorial alignments. These were independently labeled by two annotators, using the same methodology as Media Bias/Fact Check, and later discussed until full agreement. 
Final orientations are given in Tables \ref{tab:channels_upload_metrics} and \ref{table:network_summary_table} in the appendix.
\paragraph{Channels, videos, and comments collection.}

We set our collection period from March 1st, 2024, to July 14th, covering more than three months before the European elections to a week after the second round of the French legislative elections. Using the YouTube API~\cite{youtubedataAPI}, we collected channels' metadata, all videos they posted during our period of interest, and all comments posted on these videos, for a total of 43.5k videos and 7.4M comments from 700k commenters. \changed{We further collected videos from the 2017 legislative election (11th and 18th June) campaign, to evaluate the evolution of the channels. We defined a collection window of the same number of days as our main collection, ending one week after the second round of the legislative election. Because the 2017 legislative elections occurred two months after the presidential election, this window includes a portion of the 2017 presidential campaign, making it a comparably intensive election period than our main period of interest. The YouTube API imposes a limit of 20k videos when collecting videos from a channel using the upload playlist, and is not suited to collect videos older than a few months through the \textit{search} endpoint~\cite{Efstratiou_2025}. As some channels uploaded more than 20k videos since 2017, we used the \texttt{yt\_dlp}~\cite{yt-dlp} library to collect 2017 videos from this set of channels.}

\section{Methods and Definitions}\label{sec:technical}

\subsection{Shorts labeling}

Shorts are an alternative format to regular videos (RVs) on YouTube, consisting of videos under 1 minute (at the time of the collect), in a square or vertical format. They are showed to users in a dedicated tab, in which users can scroll down Shorts, without actively choosing which Short they want to see next. Following the method from \citet{Violot_Elmas_Bilogrevic_Humbert_2024}, we classified videos between Shorts and RVs by sending a request to \texttt{youtube.com/shorts/<videoid>} for each video ID of our dataset and checked if "/shorts/" was in the URL response (meaning the video was a Short) or had been removed from the response (meaning the video was not a Short).

\subsection{Graph representations}

\paragraph{Bipartite graphs and projected graphs.}

A bipartite graph $ B(U, V, E) $ is a graph where $ U $ and $ V $ are two disjoint sets of nodes and $ E $ is the set of edges such that each edge $ e \in E $ connects a node in $ U $ to a node in $ V $, but no edge exists between two nodes in $ U$ or two nodes in $V$. From $ B(U, V, E) $, we can extract the $U$-projected graph (or the $V$-projected graph). These projected graphs consist of nodes from $U$ (resp. $V$), where edges are created between nodes if they share a common neighbor from $V$ (resp. $U$). 

As part of this study, we represent each channel by its bipartite \textbf{video-commenter graph} (VCG), with edges connecting commenters to videos they commented. We further construct the \textbf{commenter-projected weighted graph} (CPWG). In this graph, commenter nodes are connected if they commented on the same video(s). Edges in the CPWG are weighted based on the number of common videos co-commented by the two commenters.

We also represent whole channels and their associated commenters by a channel-commenter bipartite graph, with edges connecting commenters to the channels whose videos they commented on. We then construct the \textbf{channel-projected weighted graph} (ChWPG), where channel nodes are connected if they share at least one commenter. The weight of each edge represents the Jaccard similarity between the commenting sets of the two channels---that  is the fraction of common commenters out of all the two channels commenters. 




\subsubsection{Augmented channel bipartite graphs.}

\begin{figure}[ht]
    \fboxsep=0pt
	\includegraphics[width=\columnwidth]{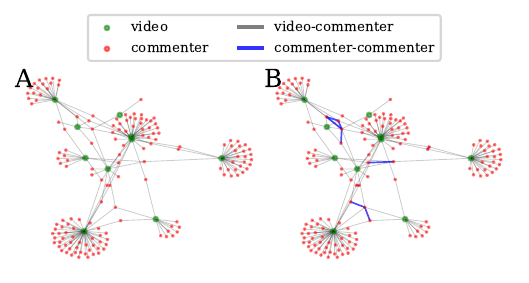}
	\caption{Example of an AVCG, using the videos and commenters from the channel ``Les Républicains". \textbf{A} shows the bipartite network with videos and commenters, and an edge when a commenter commented on a video. \textbf{B} shows the same network with added edges between commenters if they co-commented on at least two videos from the channel.}
	\label{fig:example_augmented_network}
\end{figure}

For each channel, we further combined the VCG and CPWG, using only edges with a weight $\geq 2$ to construct the \textbf{augmented video-commenter graph} (AVCG). This graph connects videos to commenters, as well as commenters who co-commented on several videos from the channel, allowing us to identify strong community patterns. Figure~\ref{fig:example_augmented_network} shows the bipartite graph of a channel and its corresponding AVCG.

\subsection{Graph measures}

Let $G=(V,E)$ be a connected graph, where $V$ is the set of vertices and $E$ is the set of edges. The \textbf{\textit{density}} of G is given by
$
D(G) = \frac{2|E|}{|V|(|V| - 1)}
$ and measures how full a graph is, that is, how many edges it has in comparison to the maximum number of edges it could have. The \textbf{\textit{transitivity}} measures the likelihood that two nodes connected to a common neighbor will also be connected together. It is given by $t = \frac{|\{triangles\}|}{|\{triplets\}|}$, where a triplet is a set of three nodes with at least two distinct edges, and a \textit{triangle} is a set of three mutually connected nodes. High transitivity indicates tightly clustered groups. The \textbf{\textit{diameter}} of $G$, denoted $diam(G)$, is given by $diam(G) = \max_{u,v \in V} d(u,v)$, where $d(u,v)$ represents the geodesic distance (i.e., the length of the shortest path) between nodes $u$ and $v$. A small diameter indicates a tightly interconnected structure.
 
 \changed{\paragraph{Densities biases.} In our experiments, for each channel, we compute the densities of the VCG,  AVCG, and CPWG. Each density measure introduces distinct biases, but together they offer a nuanced description of the channels under study. The VCG density is upper-bounded by the product between the number of videos and the number of comments and inherently favors graphs where these quantities are evenly balanced.
 The AVCG density is lower bounded by the channel's VCG density but also takes into account the proportion of commenters who co-commented on at least two videos from that channel. The difference between these two densities serves as an indicator of the commenters community’s cohesion.
 Finally, the CPWG density is biased towards channels with a low number of videos, reaching its theoretical maximum of 1 in the extreme case of a single-video channel. Because of that, the UDI party channel, which only had 10 comments, was removed from our results. }

\subsection{Politicians' Presence in News Media Videos}

Using parties' official websites, parties' lists from current elections, and journalistic news for up-to-date information, we curated a list of 1213 current politicians from all parties. 
We then used pre-trained LLMs to accomplish two tasks; \textit{mentions extraction} and \textit{interviewees extraction}, to infer which politicians were talked about and which were invited in NM channels. 

\paragraph{Mentions extraction.}  \changed{To identify mentions of politicians in video titles, we relied on a pre-trained CamemBERT model fine-tuned for named-entity recognition (NER), namely camembert-ner~\cite{camembertNER,Martin_Muller_Suarez_Dupont_Romary_Clergerie_Seddah_Sagot_2020}, available through the HuggingFace Transformers library. The model was downloaded and executed locally on a standard workstation (Apple M1 Pro, 16 GB RAM). Each video title was processed individually through the model using the HuggingFace NER pipeline with simple aggregation, which returns entities grouped by category. From the output, we retained only the entities classified as persons. According to the model’s documentation, this category achieves a precision of 0.94, a recall of 0.96, and an F1-score of 0.95, which supports the reliability of the extracted mentions.}

\paragraph{Interviewees extraction.}
\changed{To identify who was speaking in the videos, we concatenated each video title and description into a single text input using the format: \texttt{"Title: <video title>$\backslash$nDescription: <video description>"}. 
This input was then provided to an LLM with explicit instructions to extract only the names of speakers, excluding individuals who were merely mentioned (the complete prompt is available in the appendix).
We compared the extracted names to our curated list of politicians and kept only those that corresponded to political figures. 
We evaluated four models: GPT-4o~\cite{gpt4o}, GPT-4o-mini~\cite{gpt4omini}, Google’s \texttt{flan-t5-base}~\cite{text2text}, and \texttt{Mistral-Small-24B-Instruct-2501}~\cite{mistral24B}, all run with a temperature of 0 and a maximum output length of 40 tokens.
GPT models were accessed through OpenAI's API, while \texttt{flan-t5-base} and \texttt{Mistral-Small} were downloaded from HuggingFace and executed on an NVIDIA A100 GPU node. 
To compare the models, we manually annotated 200 randomly selected videos and identified the politicians interviewed. Among tested models, \texttt{Mistral-Small} performed the best, with a precision of 0.78, a recall of 0.82 and an F1-score of 0.80. Table \ref{tab:model_performances} in the appendix presents comparative performance data.}
	
\section{Results}\label{sec:results}

\subsection{Upload and Engagement Metrics (RQ1\&2)}

\begin{figure}[ht]
	\centering
	\fboxsep=0pt
	\includegraphics[]{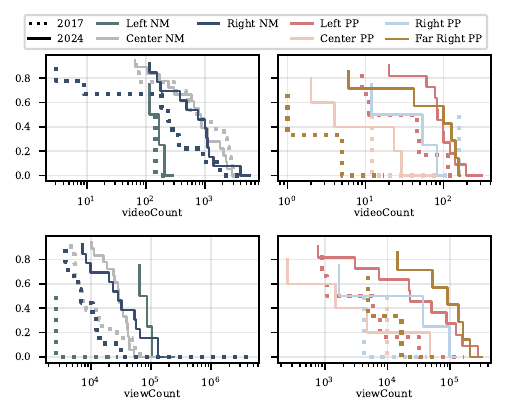}
	\caption{\changed{Empirical CCDFs of uploads and view counts, grouped by orientation. For each NM and PP orientation, we aggregated all videos from channels of that orientation and computed the distribution of uploads and view counts.}}
	\label{fig:cdf_upload_engagement_orientation}
\end{figure}

We began our analysis by examining how channels of different types and political orientations used YouTube during the 2024 elections period, and how audiences responded in terms of engagement. Table \ref{tab:summary_orientation} reports the number of NM and PP channels by orientation, along with their upload and engagement metrics. Detailed results for individual channels are provided in the appendix. \changed{To place these results in context, we also traced the evolution of uploads and views between 2017 and 2024. For each channel and election period, we calculated the total number of uploaded videos and the average number of views, and then plotted the empirical complementary cumulative distribution functions (CCDFs) of each metric by channel type and orientation (Figure \ref{fig:cdf_upload_engagement_orientation}).}

\begin{table}
	\centering
	\footnotesize
	\setlength{\tabcolsep}{3pt}
	\begin{tabular}{l|r|rr|rrr}
		& \multicolumn{1}{c|}{\multirow{3}{*}{\shortstack{\#\\chan.}}}  & \# vid. & \multirow{3}{*}{\shortstack{Shorts\\(\%)}}  & \multirow{3}{*}{\shortstack{views\\/vid.}} & \multirow{3}{*}{\shortstack{likes\\/vid.}} & \multirow{3}{*}{\shortstack{comm.\\/vid.}} \\
		& \multicolumn{1}{c|}{} & /day& & &\\
		&  \multicolumn{1}{c|}{} &  /chan. & & & \\
		\hline
		Left NM & 4 & 1.44 & 25.2 & 123k & 4.7k & 713 \\
		Center NM & \textbf{18} &  9.36 & 12.5 & 35.7k & 526 & 150 \\
		Right NM & 13 & \textbf{9.95} & 28.7 & 55.9k & 1.0k & 258 \\
		\hline
		Far Left PP & 2& 0.43 & 1.30 & 6.5k & 383 & 93.1 \\
		Left PP & 11 & 0.86 & 31.1 & 96.3k & 4.5k & 571 \\
		Center PP & 4 & 0.26 & 16.6 & 25.3k & 469 & 246 \\
		Right PP & 4  & 0.47 & 23.6 & 64.9k & 4.3k & 480 \\
		Far Right PP & 7  & 0.77 & \textbf{43.5} & \textbf{138k} & \textbf{7.6k} & \textbf{1.1k} \\
	\end{tabular}
	\caption{Collection of \textbf{channels upload metrics} (number of videos per day per channel, percentage of Shorts) and \textbf{engagement metrics} (views/video, likes/video, and comments/video), grouped by orientation. Each metric was computed for every channel and averaged over channels of the same orientation.}
	\label{tab:summary_orientation}
\end{table}

\paragraph{Upload metrics.} Table \ref{tab:summary_orientation} shows that both center and right-leaning NM channels were similarly very active, with nearly 10 videos per day on average, while left-leaning NM channels' upload activity was closer to PP channels than to other NM channels. \changed{The disparity in upload activity reflects distinct ways of using YouTube. First, when it comes to NM channels, there are two types to consider: (1) channels affiliated with TV/radio channels and (2) channels associated with the printed (online) press. The vast majority of videos posted by channels in the first category are snippets from television or radio broadcasts, while videos from channels in the second category are only intended for social media. In our dataset, centrist and right-leaning NM channels are primarily in group 1, whereas left-leaning NM channels are solely in group 2, which explains the lower uploads number.}

 \changed{Looking at the upload activity evolution in Figure \ref{fig:cdf_upload_engagement_orientation}, we can confirm that the majority of center- and right-leaning NM channels posted over 1,000 videos in 2024. While center-leaning NM channels were already active in 2017, we can see that right-leaning NM channel activity has expanded significantly since 2017.}

PP channels are less active, posting about one video every two days. Among them, left-wing and far-right channels were much more active, averaging 0.86 and 0.77 videos per day, compared to upload rates ranging from 0.26 to 0.47 for the other political orientations. \changed{This activity gap widened between 2017 and 2024 as both left-wing and far-right channels increased their production, with the far-right showing the most dramatic rise.  In contrast, the activity of right-wing PP channels decreased.}

Far-right and left-wing also post more Shorts than other orientations, especially far-right, whose channels had 43.5\% Shorts on average, and where the main party (RN) had 98.6\% Shorts. Among NM channels, the percentage of Shorts of left- and right-leaning channels is twice higher than that of center channels. 

\paragraph{Engagement metrics.} Engagement metrics analysis deserves a moderate dive into the French political landscape, to understand the diversity of audience responses among parties of the same orientation. Far-right channels lead in engagement, driven largely by the young leader of the main party (RN), J.~Bardella. His channel, launched in 2019, averages over 310k views/video. 
The other main figure of the party, presidential candidate M.~Le~Pen also has a very popular channel \changed{(107k subscribers)}, while E.~Zemmour and F.~Philippot, even more radical figures, had respectively the channels with the highest number of commenters per video \changed{(1.9k)} and the highest number of likes per video \changed{(16.4k)}. 
The left-wing group shows similarly high engagement, mainly due to the LFI party---founded in 2008 after splitting from the traditional left PS party. LFI leader J.-L.~Mélenchon has 1.1M subscribers, by far the most of any PP channels \changed{(the second highest value was 511k at the time of collect)}, and around 308k views/video on average. 
In contrast, the traditional left-wing party (PS) channel averages just 3k views per video.
Right-wing channels show relatively high engagement, with an average of 64.9k views per video. 
This is not due to the main party (LR), \changed{which only has a mean of 1.7k views per video,} but due to an alternative right party, the UPR, \changed{with a mean of 110.6k views per videos,} a party known for its active presence online, and its strong militant core~\cite{UPRmilitant}, \changed{and to a lesser degree the DLF party with 36.9k views per video}.
Far-left and centrist channels, except a relative success from president Macron's channel, show significantly lower engagement. 

\changed{Therefore, it appears that for politicians, engagement on YouTube is not necessarily tied to legacy media presence, which confirms that the platform can be an effective way to reach an audience. Channels tied to high-profile leaders (Mélenchon, Bardella) attract disproportionate engagement. However, this is also true for politicians less covered or excluded from legacy media, such as F. Asselineau and its party UPR and F. Philippot and its party LP.} We can see on Figure \ref{fig:cdf_upload_engagement_orientation} that this clear superiority of left-wing and far-right parties in terms of user engagement is a fairly recent phenomenon. In 2017, engagement results were not as distinguishably in favor of the two orientations.

Concerning NM channels, left-wing channels have by far the most views, likes, and comments per video, followed by right-wing channels, and in last, center channels with 3.5$\times$ fewer views, 9.4$\times$ fewer likes and 4.7$\times$ fewer comments per video than left-wing channels. 
\changed{As previously stated, left-wing NM channels have their content tailored for and exclusive to social media, which can explain the higher engagement they generate.} 

\paragraph{Impact of Shorts.} Shorts have been shown to generate higher engagement than RVs, both overall and within the same channel~\cite{Violot_Elmas_Bilogrevic_Humbert_2024}. To test whether this holds for French NM and PP channels, we compared the mean number of views in Shorts and RVs for each channel. We found completely diverging results among channels, from a very negative effect, Shorts generated 2 to 10$\times$ fewer views than RVs for presidential party members (center) and the PS (left-wing), but 10 to 15$\times$ more views for LFI (left-wing), RN (far-right), and UPR (right-wing) members. It appears that for already popular channels, the use of Shorts had a positive effect, whereas for less popular channels, the effect was negative. We then computed the Pearson correlation between a channel’s percentage of Shorts and its mean number of views (Shorts and RVs), finding a modest correlation of 0.59 (p = 0.001) for PP channels and none for NM channels (0.00, p = 0.99). This paints a subtle effect, where Shorts are either associated with more or fewer views than RVs within a given channel, and have a very limited effect on the overall popularity of the channel. As this result contradicts the general effect of Shorts given in~\citet{Violot_Elmas_Bilogrevic_Humbert_2024}, a further investigation of what differentiates Shorts content across channels---NM and PP---would shed light on this mixed result.\\

\changed{In the light of upload metrics, engagement metrics and usage of Shorts, we observe an interesting pattern. Far-right groups, LFI, and non-traditional right groups appear to be more active and draw more engagement than other groups. All historically faced limited visibility in legacy media and social media provided them with an opportunity to reach audiences that were otherwise hard to reach. Thus, results align with the notion of media bypass by politicians seeking unmediated visibility~\cite{Kruikemeier_Gattermann_Vliegenthart_2018}.} \changed{The media landscape has since shifted, with far-right politicians and LFI members now well established in mainstream outlets. However, their early adoption of social media, YouTube included, suggests a more flexible and effective use of the platform, shown by their activity and the disproportionate engagement they generate. In contrast, centrist actors tend to adhere to traditional media strategies. Beyond upload frequency, this is best exemplified by the use of Shorts, a social-media native format, well leveraged by the left and far-right, and overlooked by centrist channels.}

\subsection{Commenters Network Metrics (RQ2)}

\begin{figure*}[ht]
	\centering
	\includegraphics[width=\linewidth]{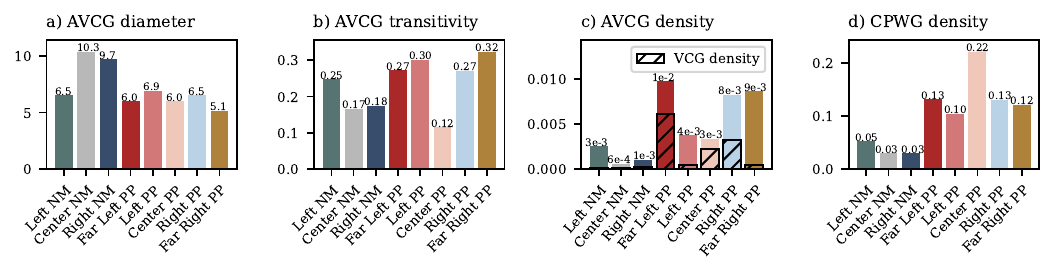}
	\caption{Network measures aggregated over channels' orientations. Each measure was computed for each channel and then averaged over channels of the same orientation.}
	\label{fig:network_measures}
\end{figure*}

We now analyze commenters networks, depending on the types of channel (NM or PP) and their orientations.  For each channel, we retrieved its bipartite video-commenter graph (VCG), computed its commenters-projected weighted graph (CPWG), and used the latter to add edges to the video-commenter graph between commenters who  co-commented on more than one video, yielding the channel's AVCG. As metrics for comparing the different channels' networks, we used the diameter, transitivity, and density of the AVCG, along with the densities of the VCG and CPWG. Results are shown in Figure~\ref{fig:network_measures}. 

From Figures~\ref{fig:network_measures}(a) and~\ref{fig:network_measures}(b), we observe that the diameter is smaller for PP channels \changed{($\mu=6.2 $, $\sigma=1.9$)} than for NM channels \changed{($\mu=9.7$, $\sigma=2.7$)}, and that the transitivity is generally higher for PP channels  \changed{($\mu=0.28 $, $\sigma=0.13$)} than NM channels \changed{($\mu=0.18$, $\sigma=0.09$)}. Left-leaning NM channels are an exception, averaging a diameter of 6.5 and a transitivity of 0.25. Centrist PP channels have low transitivity on average (0.12), with the exception of President Macron's channel, which has a transitivity of 0.31. These two values indicate that PP channels act more like small worlds~\cite{Travers_Milgram_1969} than NM channels, with more clusters and closer commenters. 

The VCG density, shown in Figure~\ref{fig:network_measures}(c) in dashed bars, is the highest for far-left and right-wing PP channels, which both have a more balanced number of videos and commenters. When examining the AVCG density, also in Figure~\ref{fig:network_measures}(c), we see that it is only 2$\times$ higher than the VCG density for these channels. In contrast, left-wing and far-right PP channels initially have low VCG densities. However, the AVCG density exhibits a substantial scaling effect---increasing by a factor of 17 for far-right channels and 8 for left-wing channels. If we remove the traditional left-wing PS party, the growth factor reaches 16 for left-wing channels. 

As explained previously, a low number of videos can explain a high CPWG density, as shown in the case of center channels which have the lowest number of videos and therefore exhibit a very high density. The rest of the PP channels groups have similar CPWG densities, systematically higher than NM channels' ones. \\

\changed{To conclude, in regards to all the network metrics, PP channels foster more strongly connected networks than NM channels, especially compared to center- and right-leaning NM channels, showing PP channels ability to build communities, whereas NM channels are more porous and diverse. 
However, there are differences among channels based on orientation. For example, left-leaning NM channels' network metrics are closer to PP channels' than to other NM channels'. Conversely, centrist and right-wing PP channels exhibit weaker commenting communities than other PP channels, with less clusters and low VCG and AVCG densities, indicating a lack of a mobilized base. 
We can tie these loose networks to the low activity of centrist and traditional left/right PP channels, shown in the previous section, but also to the likely low participation of their supporters, due to the profile of these party voters~\cite{IpsosSociologieElectorat20214} who rely more heavily on traditional media such as television and radio for political information, rather than social media~\cite{Bousquet_Figeac_Neihouser_2024}. 
These two factors cause a low participation of both politicians and their supporters on social media for centrist and traditional left/right parties.}

\subsection{Common Commenters (RQ3)}\label{sec:commoncom}

Shifting from the analysis and comparison of individual channels, we now examine how these channels might be interconnected through shared commenters. 

\subsubsection{Commenters activity.}
We begin this cross-channel analysis by evaluating the number of comments per commenter, dividing them into five groups: 1) only NM commenters, 2) only PP commenters, on same orientation channels, 3) only PP cross-orientation commenters who commented on PP channels of at least two distinct orientations, 4) Cross-type single-orientation commenters who commented on both NM and PP channels, of the same orientation, and 5) cross-type cross-orientation commenters, similar to group 4 but with PP channels of at least two distinct orientations. 

We found 67\% of only-NM commenters, with 4 comments per commenter on average; 10.7\% of only-PP single-orientation commenters with 2 comments per user; 0.6\% of only-PP cross-orientation commenters, with 5 comments per user; 14.8\% of cross-type single-orientation commenters with 18 comments per user; and 6.9\% of cross-type cross-orientation commenters with 70 comments per user. 

Focusing on commenters who commented on at least one PP video, we found that 33.0\% commented only on  left channels, 48.3\% commented only on right channels, and 15.2\%  commented on both, the remaining part consisting of commenters who also commented on center channels. We see that, while the percentage of commenters engaging with channels of two different orientations is relatively low compared to those interacting exclusively with same-orientation channels, it remains a non-negligible portion.

\begin{figure}[t]
	\centering
	\fboxsep=0pt
	\includegraphics{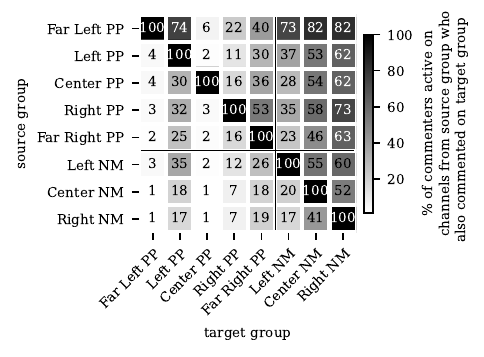}
	\caption{\changed{Percentage of commenters who commented on videos from channels in the y-axis group who also commented on videos from the x-axis group. Commenters are counted as part of a group if they commented at least once on any channel in that group.}}
	\label{fig:heatmap_common_commentors}
\end{figure}

\paragraph{Percentage of common commenters.}
Again grouping channels by type (PP vs. NM) and orientation, we examine the commenter overlap across these groups. For each group, we identify commenters who commented at least once on channels from that group and compute the percentage who also commented on other groups’ videos. Figure~\ref{fig:heatmap_common_commentors} presents these percentages in the form of a heatmap.

We confirm that those who comment on PP channels also largely comment on NM channels, as shown by the high values of NM channels columns \changed{[23-82\%]}, whereas the opposite is not true, \changed{with a majority of values $<15$\%} on PP channels columns. Right-leaning NM channels, and to a lesser degree center-leaning ones, attract particularly large portions \changed{[52-82\%]} of commenters from all orientations. \changed{Left-leaning NM channels report lower values, which is expected given the much lower volume of uploads compared to other orientations.} Among PP groups, left-wing and far-right channels receive a large share of commenters from other groups. We see that commentators from far-left channels are also extremely active on most other groups \changed{[6-82\%]}. A majority \changed{(53\%)} of commenters from right-wing channels are also commenting on far-right ones.  

\begin{figure}[ht]
	\fboxsep=0pt 
	\includegraphics[width=\columnwidth]{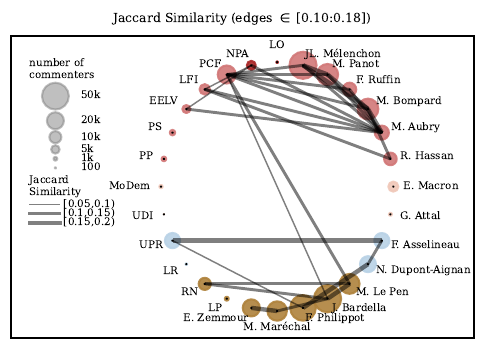}
	\caption{ChPWG of all PP channels in our dataset, with weighted edge between the 25 pairs with highest Jaccard similarities between channels' commenters.}
	\label{fig:political_channel_channel}
\end{figure}

\subsubsection{Political channels network.} We then move to the analysis of the channel-projected weighted graphs, first focusing only on PP channels. The density of the PP ChPWG was 0.977, indicating that almost all PP channels shared at least one commenter. 
Figure~\ref{fig:political_channel_channel} shows the 25 highest Jaccard similarities between channels' commenters.

We observe that channels with the same orientation are often strongly connected, except for centrist channels which do not appear in our top edges. 
One notable exception is the strong Jaccard similarity between M. Le Pen (far-right) and the PCF (left-wing). Upon investigation, we discovered several profiles of commenters. Some defend either left, either right/far-right ideas, some defend both, or some defend other parties. We translated some examples below:

\noindent\begin{tabular}{|p{3.1in}|}
	\hline
	\textbf{From a presumed left-wing commenter:}\\
	- on PCF videos: \textit{``Support all left candidates."} \\
	- on M. Le Pen videos: \textit{``L00SERS"} (untranslated) \\
	\hline
\end{tabular}
\noindent\begin{tabular}{|p{3.1in}|}
	\hline
	\textbf{From a presumed right-wing commenter:}\\
	- on PCF videos: \textit{``The left in all its splendor."} \\
	- on M. Le Pen videos: \textit{``No mercy, no forgiveness for those who don’t vote RN on July 7."} \\
	\hline
\end{tabular}
\noindent\begin{tabular}{|p{3.1in}|}
	\hline
	\textbf{From a presumed troll:}\\
	- On PCF video mentioning RN: \textit{``I express my solidarity and total support to any populist movement"} \\
	- On M. Le Pen videos: \textit{``Let’s mobilize again for the victory of the RN"} \\
	- Also on M. Le Pen videos: \textit{``Mobilize against the RN!"} \\
	\hline
\end{tabular}
\vspace{0.2cm}

\subsubsection{Political-News channels network.} We then examine the relationships between NM channels and PP~channels, using all channels to create the ChWPG. The density of the obtained graph is 0.994. Our goal was to highlight interactions between NM channels and PP channels, so we removed links connecting channels of the same type (i.e., links between two NM channels or two PP~channels).

\begin{figure}[ht]
	\centering
    \fboxsep=0pt
	\includegraphics[width=\columnwidth]{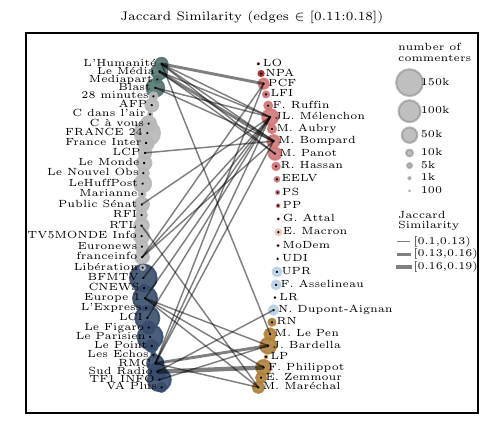}
	\caption{ChPWG of all channels showing the 30 edges with the highest Jaccard similarity. Edges between channels of the same type (NM or PP) were removed.}
	\label{fig:political_news_channel_channel}
\end{figure}

Figure~\ref{fig:political_news_channel_channel} displays the Jaccard similarity among the 30 most similar channel pairs. We observe that, in contrast with Figure~\ref{fig:political_channel_channel}, based only on PP channels, the size of NM channels has less impact on the number of strong edges. For example, while being relatively smaller in number of distinct commenters, the \textit{franceinfo} channel and the \textit{L'Humanité} channel have stronger edges than some larger channels, such as \textit{France 24} or \textit{BFMTV}. We also notice many strong edges connecting left-wing PP channels and NM channels from all political orientations, especially right-leaning NM, confirming earlier results. This effect is unrelated to channels commenters count, as illustrated by M.~Bompard’s (left) high edge count despite far fewer commenters than far-right figures like J.~Bardella or F.~Philippot.
Additionally, and similarly to results from Figure~\ref{fig:political_channel_channel}, we observe a strong edge between M.~Le~Pen and the NM channel L'Humanité which is close to the PCF party. 

\changed{Hence, what we can refer to as \textit{commenting echo chambers}---commenting mainly within a given political group/orientation---coexists with cross-posting. While commenters often stick to ideologically aligned channels, we show a non-negligible crossover between channels of opposite orientations. This supports a hybrid commenting behavior, where users comment inside partisan spaces but also interact with oppositional content.  
Our observations, confirmed by prior research~\cite{Wu_Resnick_2021}, suggest that such interactions often include disagreement or trolling. Importantly, those who engage across political lines are not typical users—they are highly active commenters, far more prolific than the average. Therefore, cross-posting behavior does not undermine the idea of echo chambers' narratives which persists for most users.
It rather supports a political agency of these users, as previously pointed~\cite{Khan_2017}, willing to devote time and effort into offering their agreement, disagreement, or nuance to the content in the videos. }
	
\subsection{News Channels Coverage of Politicians (RQ4)}\label{subsec:mentions_interviews}

Our last collection of results concerns the presence of politicians in NM channels' videos and its impact on commenters. 

\paragraph{Who's featured?}

\begin{figure}[t] 
	\centering
	\fboxsep=0pt 
	\includegraphics[width=1\linewidth]{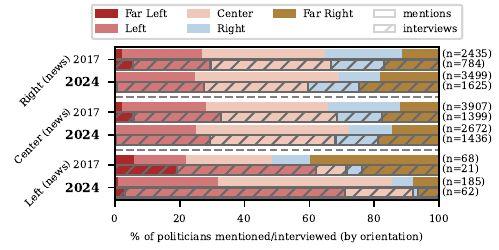}
	\caption{Political orientation of politicians mentioned or interviewed in NM videos, grouped by NM channel orientations and year, showing the evolution of politicians featured between 2017 and 2024.}
	\label{fig:fractionmentionsinterviews}
\end{figure}

We begin by investigating the distribution of political orientations among politicians mentioned or appearing on NM channels. For each NM video, we applied the Named Entity Recognition (NER) task and the interviewees extraction task, described in Section~\ref{sec:technical}, to identify politicians who were mentioned or appeared in the video, and matched them to their political orientation. Videos were then grouped by the orientation of their channel, allowing us to estimate the share of mentions or appearances for each political orientation. \changed{This procedure was repeated for both election periods, with results shown in Figure~\ref{fig:fractionmentionsinterviews}. For 2017, we excluded videos published between the two presidential rounds to avoid bias from the disproportionate visibility of the two runner-ups and their allies, whose media presence reflected first-round results rather than NM editorial choices.}

First, focusing on the 2024 results, we see strong similarities between center- and right-leaning NM. Both mainly highlight centrist politicians, followed by left-wing and far-right figures, while right-wing politicians receive less attention and the far left is absent. Centrist politicians dominate mentions across both types of channels, though their prominence decreases in interviews, especially on right-leaning NM where they are outnumbered by right and far-right figures combined. By contrast, 75\% of the guests on left-leaning media are from the left, including a few from the far left. Figures from right-wing and far-right were occasionally mentioned, but almost never featured in interviews. When that is the case, it is always from what we can call ``second-hand interviews", where extracts of interviews from other media are shared and commented. Centrist politicians are rarely interviewed but are often mentioned. 

\changed{Comparing with 2017, we first notice that right-leaning channels have doubled the number of political interviews uploaded between 2017 and 2024. Over the same period, the share of far-right mentions and interviews increased impressively, going from 12\% (resp. 17\%) to 15\% (resp. 19\%) in center NM and from 11\% (resp. 17\%) to 18\% (resp. 25\%) in right-leaning NM. }
\changed{On left NM channels, the opposite holds, with far-right politicians being overly mentioned and shown (from second-hand interviews) in 2017, by the unique active left NM channel of our dataset at the time. The share of far-right politicians then dropped between 2017 and 2024, mostly to the benefit of centrist politicians. During both election periods, politicians on the left were largely predominant. Finally, we can see the coverage of far-left politicians decreasing for left-leaning NM, and disappearing in center- and right-leaning channels in 2024.}

\changed{Therefore, it appears that there is a clear distinction between which politicians are named and which are interviewed in the videos posted by the various NM channels. Notably, far-left politicians are absent from center- and right-leaning NM channels, but the same channels do not show the same restraint with far-right politicians, who are frequently mentioned and interviewed. Conversely, left-leaning NM channels exhibit the inverse bias: they mainly mention and interview politicians from left-wing groups, and while far-right figures are often discussed, they are not invited. This dual phenomenon reveals a form of editorial gate-keeping in which channels selectively promote certain political actors while marginalizing others.}

\changed{An important nuance is that not all NM outlets operate under the same constraints. At first glance, left-leaning NM channels could appear more partisan than right-leaning ones. But in reality, right-leaning and center NM are, for most of them, radio and TV broadcasters available on the FM band and TNT (terrestrial television) and are subject to certain obligations to pluralism in exchange for access to a broad audience. Hence, the growing share of far-right politicians on right-leaning and center channels, and the disappearance of far-left ones, can have a significant impact on the general public. This contrasts with the online-only presence of left-leaning outlets which typically reach self-selected, ideologically aligned users~\cite{Bousquet_Figeac_Neihouser_2024} but enjoy more freedom in their choice of guests.}

\paragraph{Impact of politicians' presence on commenters.}

\changed{We examine how the mention or presence of politicians in an NM video changed its commenting user base. To do this, we initially split videos based on whether a politician was mentioned or not. For each NM channel, we compare the number of commenters on videos from the two groups, and found that, on average across channels, videos featuring a politician attracted 1.86$\times$ more commenters than those without.}

\begin{figure}[!t]
	\centering
	\fboxsep=0pt
	\includegraphics[width=1\linewidth]{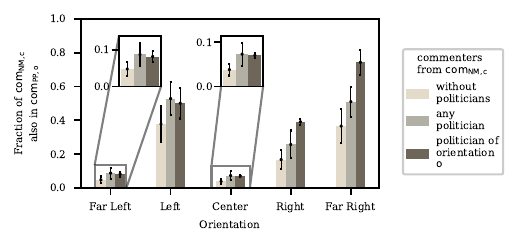}
	\caption{Impact of politician presence in NM videos on audience overlap with PP channels. For each channel and for each category $\mathsf{c}$ of NM videos (no politician, any politician, and specific orientation $\mathsf{o}$), we show the average percentage of commenters in that category ($\mathsf{com_{NM,c}}$) who also commented on PP videos ($\in \mathsf{com_{PP,o}}$). Results are shown separately for overlaps with PP channels of the same orientation and of different orientations.}
	\label{fig:comsimilarityevolutioninterviews}
\end{figure}

\changed{To better understand the origins of these commenters, we further classified the videos of each NM channel into three categories: (1) no politicians interviewed, (2) any politician interviewed, and (3) politicians from a specific political orientation interviewed. This gives us seven subsets of videos per channel: one for category (1), one for category (2) and five for category (3), one for each possible orientation of the politician being interviewed. For each video subset we extracted the corresponding set of commenters which commented on a video from that subset. We therefore obtain seven commenter sets: $\mathsf{com_{NM,no\,polit}}$ for group (1), $\mathsf{com_{NM,any\,polit}}$ for group (2), and, for each orientation $\mathsf{o}$ in group (3), $\mathsf{com_{NM,polit\,from\,o}}$.}

\changed{We also grouped PP videos by their political orientation and, for each orientation $\mathsf{o}$, extracted the complete set of commenters who engaged with those videos, denoted $\mathsf{com_{PP,o}}$. For each NM channel and each PP orientation, we then computed the percentage of commenters from each set $\mathsf{com_{NM,c}}$, where $\mathsf{c \in [no\,polit, any\,polit,polit\,from\,o]}$, that also belonged to $\mathsf{com_{PP,o}}$. In particular, we focused on the percentage of commenters from sets $\mathsf{com_{NM,o}}$, who also commented videos of politicians from the same orientation. Finally, we averaged these percentages over all NM channels and obtain the results shown in Figure~\ref{fig:comsimilarityevolutioninterviews}.}

First, looking at the effect of politicians' presence without taking their orientation into account, we observe that a politician's presence in a video increases a lot the fraction of commenters who also commented on PP videos, of any orientation. \changed{For example, for NM videos without politicians, on average, 4.8\% of commenters also commented on videos from far-left PP channels, but for NM videos with a politician this percentage increases to 8.7\%.} \changed{This holds when looking at the specific orientation of the politician being interviewed, the percentage of commenters also in $\mathsf{com_{PP,orientation^i}}$ increases with the presence of a politician with the corresponding orientation. For far-left, left, and center, the orientation of the politician does not cause a big difference in the percentage of commenters also in $\mathsf{com_{PP,orientation^i}}$. For right and far-right the percentage of commenters also in $\mathsf{com_{PP,orientation^i}}$ increases significantly when the politician interviewed is from their corresponding orientation, jumping from 25\% to 40\% for right-wing politicians compared to other orientations and from 51\% to 74\% for far-right politicians.}

\changed{Therefore, politicians' presence acts as a bridge between PP and NM channels. When an NM channel video features a politician, the percentage of its commenters who also comments on videos from politicians of the same orientation as the one shown in the video increases considerably, especially for right and far-right politicians. Additionally, videos featuring a politician of any orientation increase the presence of commenters from all groups.  This, together with the previously observed cross-commenting activity, supports the theory of commenters as politically invested actors, going beyond PP channels to support or contradict politicians, and following them into NM channels to express their opinions.}

\section{Conclusion}\label{sec:conclusion}

In this article, we examined the different ways in which political actors and news media in France used YouTube during a highly active electoral period. We found that activity and engagement differ significantly among orientations and across NM and PP channels. During these campaign periods, left-wing and far-right channels were far more active and engaging than centrist and traditional right-wing channels. They also made a higher usage of Shorts. Left-leaning NM channels were less active but received far higher engagement than center- and right-leaning ones. One striking result is the conditional impact of Shorts which led to higher engagement only for already popular PP channels. Diving deeper into Shorts dynamics and their impact on political communication is a promising direction for future research. 

Regarding commenting patterns, PP channels had tighter, more coherent commenting communities than NM channels, especially on the left and far-right. Centrist channels showed weaker communities. While commenters often remain inside ideologically aligned spaces, cross-orientation commenters were present and prolific, suggesting motivated political engagement, not random exposure.

We also showed that NM channels favor politicians with similar political leanings in their coverage and interviews, \changed{a phenomenon that intensified over the years. Left-leaning channels rarely featured far-right figures and mentioned them less and less, while center- and right-leaning channels stopped featuring far-left politicians and increasingly interviewed and mentioned far-right politicians.} In general, centrist politicians were overrepresented, at least in mentions, across all NM orientations. Regarding commenting behavior, we found that the presence of politicians in NM channels' videos significantly increased commenters overlap between PP and NM channels, not only between aligned channels but all across the political spectrum. 

\changed{As a final remark, we notice that far-right politicians not only make an effective use of YouTube in their uploading habits as well as in audience response, they also benefit from a growing share of NM videos mentioning or featuring them, from right-leaning and centrist channels. Obviously, we cannot explain their victory to the French European election to their YouTube presence only, but, considering the growing number of potential French voters who use YouTube as a source of political information, its effect should not be underestimated either.
Additionally, for TV and radio NM, YouTube is a way to re-publish their content, after a broadcast on their medium of origin, thereby doubling the exposure of invited politicians and extending their reach to new audiences~\cite{IpsosSociologieElectorat20214}. Legacy media have been increasingly criticized for mainstreaming far-right ideologies~\cite{plottu2024pop}, and YouTube appears to function as yet another vector through which these narratives are reinforced and normalized within the public sphere. }

\vspace{0.1cm}
\textit{Limitations}. This study deliberately centers on the French political-media ecosystem to move beyond the predominantly US-centric focus of prior research. While offering diversity, our findings are similarly context-specific. We focused on YouTube, but a cross-platform analysis, including social media targeting different demographics such as X, Facebook, and TikTok, could provide a more holistic view of how politicians and news media engage online. Finally, the political orientation of NM channels was in part manually annotated, based on external references and annotator judgment, which introduces a degree of subjectivity.



\bibliographystyle{aaai25}
\bibliography{aaai22}

\newpage

\section*{Appendix}

\subsection*{A - Interviewee Extraction Task}
\paragraph{Prompt for interviewee extraction from a concatenation of a video's title and description.}
\mbox{}\\
\texttt{'''\\}
\texttt{Instruction:\\}
\texttt{Extract only the full personal names of people who are invited in the video.\\}
\texttt{Exclude titles, roles, affiliations, and political parties.\\}
\texttt{Exclude people who are only mentioned but not invited or speaking.
	If no speakers, return an empty list.\\}

\texttt{Output format (strict):\\}
\texttt{{"Invited": ["Full Name 1", "Full Name 2"]}\\}

\texttt{You must output only valid JSON, no explanations, no text outside the JSON\\}

\texttt{Examples:\\}

\noindent-\texttt{Input: Titre: "Interview exclusive avec Emmanuel Macron et Jean Dupont"\\}
\texttt{Description : "Le président Emmanuel Macron s’entretient avec Jean Dupont sur les enjeux actuels."\\}
-\texttt{Output:{"Invited": ["Emmanuel Macron", "Jean Dupont"]}\\}

\noindent-\texttt{Input: Titre: "Hommage à Simone Veil"\\}
\texttt{Description: "Le président a évoqué Simone Veil dans son discours."\\}
-\texttt{Output:{"Invited": []}\\}

\texttt{Now analyze the following:\\}
\texttt{'''}
\paragraph{Models' Performances.}
\mbox{}
\begin{table}[th]
	\centering
	\begin{tabular}{c|c|c|c}
		model & precision & recall & f1-score\\
		\hline
		gpt-4o & \textbf{0.94} & 0.63 & 0.75 \\
		gpt-4o-mini & 0.68 & 0.88 & 0.77 \\
		flan-t5-base & 0.72 & \textbf{0.89} & 0.79 \\
		mistral-24B & 0.78 & 0.82 & \textbf{0.8}\\
	\end{tabular}
	\caption{Performances of each model on the extraction of interviewee task. Measures with the highest score were put in bold.}
	\label{tab:model_performances}
\end{table}

\subsection*{B - Collection of channels along with their individual upload metrics, engagement metrics and network metrics.}

See Table \ref{tab:channels_upload_metrics} for upload and engagement metrics, and Table \ref{table:network_summary_table} for network metrics.

\renewcommand{\arraystretch}{0.95}
		      
\begin{table*}[ht]
	\centering
	\begin{tabular}{|l|l|rrr|rrr|}
		
		\hline
		\multirow{3}{*}{Orientation} & \multirow{3}{*}{Channel Title} & \multicolumn{6}{c|}{Upload and Engagement Metrics} \\
		\cline{3-8}
		& & \multirow{2}{*}{subs} & \multirow{2}{*}{\# videos} & \multirow{2}{*}{\% Short} & mean & mean & mean  \\
		& & & &  & views & likes &  comments \\
\hline
\multirow{4}{*}{Left (NM)} &  L'Humanité &  183k &  208 &  68.8 &  64.2k &  3.0k &  474 \\
& \g Le Média & \g 1.1M & \g 286 & \g 0.0 & \g 103k & \g 3.6k & \g 737 \\
&  Mediapart &  800k &  115 &  29.6 &  85.7k &  2.3k &  315 \\
& \g Blast & \g 1.1M & \g 167 & \g 2.4 & \g \textbf{238k} & \g \textbf{9.9k} & \g \textbf{1.3k} \\
\hline
\multirow{18}{*}{Center (NM)} &  28 minutes &  513k &  241 &  0.4 &  29.2k &  506 &  146 \\
& \g AFP & \g 1.6M & \g 1797 & \g 13.4 & \g 11.3k & \g 116 & \g 51.7 \\
&  C dans l'air &  769k &  588 &  0.0 &  41.0k &  328 &  81.6 \\
& \g C à vous & \g 929k & \g 1103 & \g 6.3 & \g 34.6k & \g 333 & \g 124 \\
&  FRANCE 24 &  \textbf{6.7M} &  4533 &  1.4 &  39.5k &  315 &  113 \\
& \g France Inter & \g 1.2M & \g 2061 & \g 1.8 & \g 31.8k & \g 657 & \g 73.7 \\
&  LCP &  259k &  647 &  4.9 &  29.5k &  289 &  135 \\
& \g Le Monde & \g 1.8M & \g 170 & \g 34.1 & \g 124k & \g 2.2k & \g 698 \\
&  Le Nouvel Obs &  400k &  300 &  6.3 &  28.3k &  293 &  211 \\
& \g LeHuffPost & \g 1.4M & \g 862 & \g 4.1 & \g 51.8k & \g 635 & \g 241 \\
&  Marianne &  148k &  65 &  1.5 &  43.9k &  1.6k &  297 \\
& \g Public Sénat & \g 508k & \g 1123 & \g 10.3 & \g 18.4k & \g 193 & \g 109 \\
&  RFI &  1.6M &  778 &  3.2 &  43.1k &  276 &  56.5 \\
& \g RTL & \g 842k & \g 2375 & \g 43.7 & \g 41.7k & \g 615 & \g 60.4 \\
&  TV5MONDE Info &  1.9M &  2864 &  1.5 &  23.8k &  181 &  46.7 \\
& \g EuroNM & \g 2.1M & \g 2263 & \g 0.0 & \g 10.4k & \g 91.2 & \g 56.5 \\
&  franceinfo &  587k &  871 &  21.6 &  16.2k &  267 &  127 \\
& \g Libération & \g 55.4k & \g 104 & \g 70.2 & \g 24.7k & \g 591 & \g 63.9 \\
\hline
\multirow{13}{*}{Right (NM)} &  BFMTV &  2.0M &  1166 &  41.0 &  65.9k &  989 &  469 \\
& \g CNEWS & \g 491k & \g 379 & \g \textbf{72.8} & \g 23.5k & \g 592 & \g 54.7 \\
&  Europe 1 &  1.7M &  \textbf{5680} &  3.5 &  20.1k &  293 &  111 \\
& \g L'Express & \g 152k & \g 178 & \g 66.9 & \g 7.3k & \g 98.2 & \g 21.3 \\
&  LCI &  1.4M &  1006 &  11.5 &  131k &  1.6k &  891 \\
& \g Le Figaro & \g 694k & \g 759 & \g 22.1 & \g 32.6k & \g 557 & \g 175 \\
&  Le Parisien &  1.5M &  1384 &  28.7 &  159k &  3.3k &  354 \\
& \g Le Point & \g 292k & \g 1097 & \g 1.4 & \g 9.9k & \g 61.3 & \g 92.7 \\
&  Les Echos &  244k &  115 &  20.9 &  28.5k &  345 &  69.3 \\
& \g RMC & \g 541k & \g 487 & \g 10.5 & \g 53.5k & \g 255 & \g 354 \\
&  Sud Radio &  974k &  3995 &  2.1 &  8.3k &  267 &  72.9 \\
& \g TF1 INFO & \g 780k & \g 1063 & \g 24.7 & \g 130k & \g 2.4k & \g 352 \\
&  VA Plus &  449k &  154 &  66.9 &  56.6k &  2.7k &  332 \\
\hline
\hline
\multirow{2}{*}{Far Left (PP)} &  LO &  3.2k &  39 &  0.0 &  946 &  41.1 &  15.0 \\
& \g NPA & \g 29.1k & \g 77 & \g 2.6 & \g 12.0k & \g 725 & \g 171 \\
\hline
\multirow{11}{*}{Left (PP)} &  PCF &  47.4k &  \textbf{304} &  49.3 &  50.5k &  1.8k &  218 \\
& \g LFI & \g 117k & \g 99 & \g 8.1 & \g 22.1k & \g 1.4k & \g 238 \\
&  F. Ruffin (EcoS) &  258k &  60 &  68.3 &  87.4k &  4.5k &  460 \\
& \g JL. Mélenchon (LFI) & \g \textbf{1.1M} & \g 140 & \g 50.7 & \g 307k & \g 13.7k & \g 1.2k \\
&  M. Aubry (LFI) &  26.9k &  120 &  9.2 &  22.8k &  1.3k &  495 \\
& \g M. Bompard (LFI) & \g 89.4k & \g 84 & \g 29.8 & \g 157k & \g 7.6k & \g 1.1k \\
&  M. Panot (LFI) &  237k &  78 &  24.4 &  123k &  6.5k &  1k \\
& \g R. Hassan (LFI) & \g 26.1k & \g 20 & \g 75.0 & \g 279k & \g 12.9k & \g 1.4k \\
&  EELV &  7.2k &  98 &  1.0 &  7.3k &  230 &  106 \\
& \g PS & \g 13.3k & \g 82 & \g 13.4 & \g 3.0k & \g 64.8 & \g 53.0 \\
&  PP &  2.3k &  193 &  12.4 &  781 &  12.9 &  13.4 \\
\hline
\multirow{4}{*}{Center (PP)} & \g G. Attal (RE) & \g 2.9k & \g 4 & \g 25.0 & \g 52.0k & \g 501 & \g 320 \\
&  E. Macron (RE) &  339k &  29 &  41.4 &  47.4k &  1.4k &  651 \\
& \g MoDem & \g 8.9k & \g 84 & \g 0.0 & \g 1.5k & \g 12.3 & \g 12.8 \\
&  UDI &  483 &  23 &  0.0 &  254 &  3.61 &  0.48 \\
\hline
\multirow{4}{*}{Right (PP)} & \g UPR & \g 455k & \g 54 & \g 1.9 & \g 97.1k & \g 6.5k & \g 887 \\
&  F. Asselineau (UPR) &  53.1k &  82 &  26.8 &  124k &  5.6k &  477 \\
& \g LR & \g 24.3k & \g 12 & \g 0.0 & \g 1.7k & \g 58.5 & \g 25.2 \\
&  N. Dupont-Aignan (DLF) &  165k &  105 &  65.7 &  36.9k &  4.9k &  531 \\
\hline
\multirow{7}{*}{Far Right (PP)} & \g RN & \g 49.3k & \g 143 & \g \textbf{98.6} & \g 52.4k & \g 3.0k & \g 217 \\
&  M. Le Pen (RN) &  107k &  126 &  81.0 &  206k &  8.6k &  774 \\
& \g J. Bardella (RN) & \g 96.1k & \g 159 & \g 61.6 & \g \textbf{312k} & \g 12.6k & \g 1.2k \\
&  LP &  25.8k &  6 &  0.0 &  14.3k &  2.4k &  466 \\
& \g F. Philippot (LP) & \g 511k & \g 155 & \g 2.6 & \g 134k & \g \textbf{16.4k} & \g 1.8k \\
&  E. Zemmour (REC) &  480k &  42 &  28.6 &  160k &  6.1k &  \textbf{1.9k} \\
& \g M. Maréchal (REC-RN) & \g 95.0k & \g 100 & \g 32.0 & \g 89.9k & \g 4.0k & \g 1.1k \\
\hline
	\end{tabular}
	\caption{YouTube Channel Statistics.}
	\label{tab:channels_upload_metrics}
\end{table*}

\begin{table*}
	\centering
	\begin{tabular}{|l|l|rrr|rrr|rr|}
		\hline
		
		\multirow{3}{*}{Orientation} & \multirow{3}{*}{Channel Title}  &   \multicolumn{3}{|c|}{VCG graph} & \multicolumn{3}{c|}{AVCG} &
		\multicolumn{2}{c|}{CPWG} \\
		\cline{3-10}
	 &  & \# com-  & \multirow{2}{*}{dens.}  & dens. &   \multirow{2}{*}{dens.} & transi- & diam- & \multicolumn{1}{c}{\#}  &  \multirow{2}{*}{dens.} \\
	 &  & ponents & & norm & & tivity  & eter & edges &  \\

\hline
\multirow{4}{*}{Left (NM)} &  L'Humanité &  1 &  1.27e-04 &  9.27e-03 &  \textbf{4.20e-03} &  0.36 &  8 &  27.3M &  0.061 \\
& \g Le Média & \g 1 & \g 9.88e-05 & \g 8.73e-03 & \g 2.65e-03 & \g 0.24 & \g 6 & \g 50.3M & \g 0.040 \\
&  Mediapart &  1 &  2.52e-04 &  1.83e-02 &  8.38e-04 &  0.17 &  6 &  3.3M &  0.065 \\
& \g Blast & \g 1 & \g 6.50e-05 & \g 1.26e-02 & \g 2.54e-03 & \g 0.22 & \g 6 & \g 96.3M & \g 0.046 \\
\hline
\multirow{18}{*}{Center (NM)} &  28 minutes &  3 &  1.94e-04 &  5.69e-03 &  4.79e-04 &  0.14 &  9 &  2.9M &  0.031 \\
& \g AFP & \g 116 & \g 9.56e-05 & \g 1.20e-03 & \g 2.32e-04 & \g 0.09 & \g 14 & \g 5.3M & \g 0.009 \\
&  C dans l'air &  3 &  5.85e-04 &  1.52e-02 &  6.91e-04 &  0.02 &  12 &  401k &  0.059 \\
& \g C à vous & \g 19 & \g 7.47e-05 & \g 1.70e-03 & \g 4.79e-04 & \g 0.18 & \g 11 & \g 22.9M & \g 0.022 \\
&  FRANCE 24 &  52 &  3.84e-05 &  6.13e-04 &  2.80e-04 &  0.13 &  12 &  59.5M &  0.007 \\
& \g France Inter & \g 51 & \g 7.93e-05 & \g 1.06e-03 & \g 7.98e-04 & \g 0.27 & \g 10 & \g 26.6M & \g 0.023 \\
&  LCP &  19 &  9.36e-05 &  2.45e-03 &  7.66e-04 &  \textbf{0.46} &  10 &  19.3M &  0.039 \\
& \g Le Monde & \g 1 & \g 6.86e-05 & \g 8.80e-03 & \g 7.54e-04 & \g 0.21 & \g 6 & \g 39.6M & \g 0.041 \\
&  Le Nouvel Obs &  13 &  1.09e-04 &  5.00e-03 &  3.81e-04 &  0.14 &  10 &  9.5M &  0.029 \\
& \g LeHuffPost & \g 3 & \g 6.91e-05 & \g 2.46e-03 & \g 5.86e-04 & \g 0.15 & \g 8 & \g 31.6M & \g 0.018 \\
&  Marianne &  1 &  2.91e-04 &  \textbf{2.12e-02} &  2.10e-03 &  0.28 &  6 &  5.2M &  \textbf{0.115} \\
& \g Public Sénat & \g 8 & \g 8.53e-05 & \g 1.65e-03 & \g 3.61e-04 & \g 0.13 & \g 10 & \g 17.1M & \g 0.019 \\
&  RFI &  75 &  1.11e-04 &  2.06e-03 &  2.18e-04 &  0.12 &  12 &  3.5M &  0.013 \\
& \g RTL & \g 73 & \g 9.05e-05 & \g 9.88e-04 & \g 4.14e-04 & \g 0.16 & \g 12 & \g 14.4M & \g 0.016 \\
&  TV5MONDE Info &  \textbf{169} &  7.23e-05 &  7.97e-04 &  2.19e-04 &  0.12 &  12 &  8.6M &  0.008 \\
& \g Euronews & \g 35 & \g 1.47e-04 & \g 1.24e-03 & \g 7.06e-04 & \g 0.14 & \g 11 & \g 6.2M & \g 0.014 \\
&  franceinfo &  11 &  8.64e-05 &  2.02e-03 &  3.89e-04 &  0.15 &  8 &  12.7M &  0.019 \\
& \g Libération & \g 8 & \g 5.05e-04 & \g 1.17e-02 & \g 7.32e-04 & \g 0.10 & \g 13 & \g 590k & \g 0.064 \\
\hline
\multirow{13}{*}{Right (NM)} &  BFMTV &  4 &  2.87e-05 &  1.85e-03 &  1.32e-03 &  0.30 &  10 &  \textbf{412M} &  0.037 \\
& \g CNEWS & \g 19 & \g 2.57e-04 & \g 3.86e-03 & \g 5.70e-04 & \g 0.21 & \g 10 & \g 2.1M & \g 0.046 \\
&  Europe 1 &  94 &  4.78e-05 &  5.78e-04 &  9.88e-04 &  0.19 &  12 &  138M &  0.016 \\
& \g L'Express & \g 35 & \g \textbf{8.85e-04} & \g 8.40e-03 & \g 9.75e-04 & \g 0.01 & \g \textbf{18} & \g 117k & \g 0.050 \\
&  LCI &  2 &  6.63e-05 &  4.20e-03 &  3.01e-03 &  0.28 &  6 &  195M &  0.027 \\
& \g Le Figaro & \g 3 & \g 9.41e-05 & \g 4.70e-03 & \g 4.28e-04 & \g 0.19 & \g 10 & \g 14.3M & \g 0.030 \\
&  Le Parisien &  4 &  3.69e-05 &  1.75e-03 &  4.42e-04 &  0.13 &  8 &  113M &  0.014 \\
& \g Le Point & \g 39 & \g 1.37e-04 & \g 2.13e-03 & \g 7.48e-04 & \g 0.16 & \g 11 & \g 7.4M & \g 0.018 \\
&  Les Echos &  12 &  4.85e-04 &  1.08e-02 &  5.35e-04 &  0.01 &  10 &  411k &  0.042 \\
& \g RMC & \g 1 & \g 6.90e-05 & \g 3.95e-03 & \g 1.03e-03 & \g 0.25 & \g 7 & \g 54.3M & \g 0.036 \\
&  Sud Radio &  63 &  9.07e-05 &  9.32e-04 &  1.42e-03 &  0.21 &  9 &  47.3M &  0.025 \\
& \g TF1 INFO & \g 3 & \g 3.93e-05 & \g 2.05e-03 & \g 5.41e-04 & \g 0.17 & \g 9 & \g 106M & \g 0.018 \\
&  VA Plus &  1 &  2.16e-04 &  1.19e-02 &  1.39e-03 &  0.17 &  6 &  5.0M &  0.036 \\
\hline
\multirow{2}{*}{Far Left (PP)} &  LO &  1 &  1.16e-02 &  4.50e-02 &  1.45e-02 &  0.17 &  6 &  2.7k &  0.161 \\
& \g NPA & \g 1 & \g 5.54e-04 & \g 2.05e-02 & \g 4.96e-03 & \g 0.38 & \g 6 & \g 1.5M & \g 0.103 \\
\hline
\multirow{11}{*}{Left (PP)} &  PCF &  1 &  1.50e-04 &  5.78e-03 &  1.36e-03 &  0.31 &  8 &  14.2M &  0.056 \\
& \g LFI & \g 1 & \g 4.43e-04 & \g 1.72e-02 & \g 3.11e-03 & \g 0.39 & \g 6 & \g 2.3M & \g 0.080 \\
&  F. Ruffin &  1 &  2.51e-04 &  2.59e-02 &  2.38e-03 &  0.29 &  6 &  5.7M &  0.076 \\
& \g JL. Mélenchon & \g 1 & \g 8.14e-05 & \g 1.51e-02 & \g 3.15e-03 & \g 0.27 & \g 6 & \g 62.3M & \g 0.048 \\
&  M. Aubry &  1 &  3.41e-04 &  2.01e-02 &  7.72e-03 &  0.32 &  6 &  8.5M &  0.091 \\
& \g M. Bompard & \g 1 & \g 1.41e-04 & \g 2.53e-02 & \g 6.86e-03 & \g 0.32 & \g 6 & \g 38.0M & \g 0.085 \\
&  M. Panot &  1 &  1.07e-04 &  2.12e-02 &  3.27e-03 &  0.37 &  6 &  64.3M &  0.137 \\
& \g R. Hassan & \g 1 & \g 2.40e-04 & \g 8.78e-02 & \g 5.43e-03 & \g \textbf{0.54} & \g 4 & \g 15.3M & \g 0.256 \\
&  EELV &  3 &  5.91e-04 &  1.63e-02 &  4.28e-03 &  0.35 &  8 &  1.8M &  0.149 \\
& \g PS & \g 4 & \g 1.26e-03 & \g 1.80e-02 & \g 2.29e-03 & \g 0.11 & \g 8 & \g 291k & \g 0.141 \\
&  PP &  \textbf{14} &  1.48e-03 &  8.27e-03 &  1.73e-03 &  0.06 &  \textbf{12} &  28.3k &  0.027 \\
\hline
\multirow{4}{*}{Center (PP)} & \g G. Attal & \g 1 & \g 3.48e-03 & \g 2.57e-01 & \g 3.67e-03 & \g 0.00 & \g 4 & \g 69.7k & \g 0.408 \\
&  E. Macron &  1 &  3.97e-04 &  4.85e-02 &  3.08e-03 &  0.31 &  4 &  3.7M &  0.150 \\
& \g MoDem & \g 9 & \g 2.90e-03 & \g 2.10e-02 & \g 3.41e-03 & \g 0.03 & \g 10 & \g 25.0k & \g 0.105 \\
&  UDI &  3 &  \textbf{1.36e-01} &  \textbf{3.33e-01} &  \textbf{3.33e-01} &  0 &  2 &  10.0 &  \textbf{1} \\
\hline
\multirow{4}{*}{Right (PP)} & \g UPR & \g 1 & \g 2.52e-04 & \g 3.72e-02 & \g 9.26e-03 & \g 0.34 & \g 6 & \g 19.0M & \g 0.151 \\
&  F. Asselineau &  1 &  2.12e-04 &  2.05e-02 &  3.27e-03 &  0.38 &  6 &  10.3M &  0.084 \\
& \g LR & \g 2 & \g 1.23e-02 & \g 9.12e-02 & \g 1.34e-02 & \g 0.03 & \g 8 & \g 2.6k & \g 0.192 \\
&  N. Dupont-Aignan &  1 &  2.45e-04 &  2.09e-02 &  7.17e-03 &  0.33 &  6 &  14.8M &  0.096 \\
\hline
\multirow{7}{*}{Far Right (PP)} & \g RN & \g 1 & \g 4.00e-04 & \g 1.46e-02 & \g 3.50e-03 & \g 0.24 & \g 6 & \g 3.2M & \g 0.063 \\
&  M. Le Pen &  1 &  1.58e-04 &  1.73e-02 &  4.69e-03 &  0.27 &  6 &  31.3M &  0.084 \\
& \g J. Bardella & \g 1 & \g 9.06e-05 & \g 1.55e-02 & \g 4.04e-03 & \g 0.25 & \g 6 & \g 71.0M & \g 0.054 \\
&  LP &  1 &  2.35e-03 &  2.52e-01 &  1.34e-02 &  0.49 &  4 &  203k &  0.359 \\
& \g F. Philippot & \g 1 & \g 1.81e-04 & \g 2.69e-02 & \g 1.81e-02 & \g 0.34 & \g 4 & \g \textbf{101M} & \g 0.097 \\
&  E. Zemmour &  1 &  1.91e-04 &  4.62e-02 &  9.33e-03 &  0.35 &  6 &  22.5M &  0.110 \\
& \g M. Maréchal & \g 1 & \g 1.81e-04 & \g 2.37e-02 & \g 7.65e-03 & \g 0.32 & \g 4 & \g 28.2M & \g 0.083 \\
\hline
	\end{tabular}
	\caption{Channels Networks Summary Table.}
	\label{table:network_summary_table}
\end{table*}

\end{document}